# Existence of electronic dipole lattice on $CuO_2$ plane of cuprate superconductor

Dong hang

Institute of Applied Physics and Computational Mathematics, P. O. Box 8009-11, Beijing 100088, China

[Abstract]

The existing theory about the cuprate superconductor bases on band theory, almost all neglects the position difference of Cu-O ions on $CuO_2$ plane, and assumes that the charge on ions are averagely distributed. Considering that the nature of conduction is the moving of electrons, carrier means electron in this article. The charge opposite the carriers distributes on fixed $CuO_2$ ions, and the position difference of Cu-O ions slightly violates average electricity distribution, so the concept of hole is not suitable. It is proved that, near the optimal doping level, since the density of ions is relatively low, the violations can be treated as a perturbation, and the ion lattice can be seen as a combination of regular lattice and dipole lattice. Considering about that the character of cuprate superconductor is mainly dominated by the dipole lattice, the pairing mechanism, the d-wave symmetry, and the linear temperature dependence of normal state resistivity is tersely interpreted.

## 1 Introduction

Although the mechanism of cuprate superconductor remains unsolved, much related knowledge is confirmed [1-5]: superconductivity mainly occurs on the $CuO_2$ plane, d-wave symmetry, relatively short coherence length, and linear temperature dependence of normal state resistivity, pseudogap, and the like. Some new research reveals that, Cooper pairs may occur in real space [6], purely magnetic or electronic pairing mechanism [7], and so on.

The enormous research on pairing mechanism of cuprate superconductor almost all imply a common hypothesis that, on $CuO_2$ plane, the charge on Cu-O ions is opposite of that on carriers, and is homogenously distributed [8,9]. Some theories consider about the intrinsic inhomogeneous charge distribution of cuprate superconductor [10].

Differing from the existing theories, I think that the inhomogeneous ion charge distribution comes from the character of $CuO_2$ plane itself. Since the nature of conduction is the moving of electrons, carrier means electron in this article. On the $CuO_2$ plane, there exist free electrons and four kinds of ions: $Cu^{2+}$, $Cu^{1+}$, $O^{2-}$, $O^{1-}$. When $Cu^{1+}$ and $O^{2-}$ lost electron they become $Cu^{2+}$ and $O^{1-}$, and when $Cu^{2+}$ and $O^{1-}$ obtain electron they become $Cu^{1+}$ and $O^{2-}$, electron and ions keep some kind of balance on $CuO_2$ plane.

$Cu^{1+}$ and $O^{2-}$ can't attract electron anymore so they can be treated as electrically neutral, and the number of $Cu^{1+}$ and $O^{2-}$ is determined by the electronegative character of $CuO_2$ plane; $Cu^{2+}$ and $O^{1-}$ can be seen as cation, the total number of $Cu^{2+}$ and $O^{1-}$ is equal to the number of carriers, and they form the ion lattice. The charge opposite of carriers distributes on Cu-O ions with fixed position, and the position difference of Cu-O ions will slightly violate homogenous electricity distribution. The concept of hole neglects the position difference of Cu-O ions, is not suitable for $CuO_2$ plane. Around optimum doping, the concentration of ions is relatively low, and the violation can be treated as perturbation.

In this article, I study this hypothesis with theoretical analysis and experimental phenomena explanation, and prove that the ion lattice of $CuO_2$ plane can be seen as a combination of regular lattice and electronic dipole lattice around optimum doping。 The interactivity between regular lattice and carriers acts like normal metal, forms the ground state; and the electronic dipole lattice introduce additional energy to the $CuO_2$ plane, dominates the character of cuprate superconductor.

**2 Theoretical analyses**

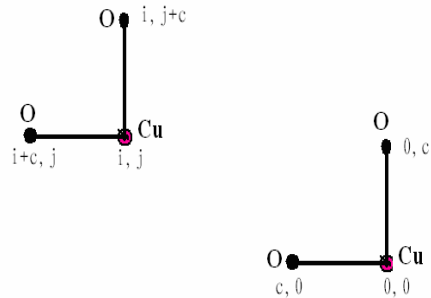

**Fig1. Illustration of ion units**

Considering the ion lattice on $CuO_2$ plane shown in Fig1, choose the $CuO_2$ as a unit of ion lattice. When the bond length of Cu-O is $b$, the distance between nearest units can be approximately written as $a_0 = b/\sqrt{x_0}$, where $x_0$ is carrier concentration. To calculate the lattice energy that created from the interactivity between an ion and the ion lattice, take $a_0$ as 1, and bond length as $c$, so $c = \sqrt{x_0}$. For cuprate superconductor, around optimum doping, $x_0$ is about 0.15, $c$ can be treated as perturbation. Defining the occupying probability on Cu ions and O ions as $P_{Cu}$ and $P_O$ respectively, then $P_{Cu}+2P_O=1$, means that one unit has only one charged ion. For simple, the coulomb screening is considered with $\varepsilon=\varepsilon_{eff}\varepsilon_0$, where $\varepsilon_0$ is electric constant, and $\varepsilon_{eff}$ presents coulomb screening. So the lattice energy is:

$$E=\frac{e^2}{4\pi\varepsilon}\sum_{i,j}\left\{\begin{array}{l}\frac{P_{Cu}^2}{\sqrt{i^2+j^2}}+\frac{2P_O^2}{\sqrt{i^2+j^2}}+P_O^2\left[\frac{1}{\sqrt{(i+c)^2+(j-c)^2}}+\frac{1}{\sqrt{(i-c)^2+(j+c)^2}}\right] \\ +P_{Cu}P_O\left[\frac{1}{\sqrt{(i+c)^2+j^2}}+\frac{1}{\sqrt{(i-c)^2+j^2}}+\frac{1}{\sqrt{(j+c)^2+i^2}}+\frac{1}{\sqrt{(j-c)^2+i^2}}\right]\end{array}\right\} \quad (1)$$

When only keep $c^2$ term, one have

$$\frac{1}{\sqrt{(i+c)^2+j^2}}=\frac{1}{\sqrt{r_{ij}^2*(1+\frac{2ic+c^2}{r_{ij}^2})}}=\frac{1}{r_{ij}}\left(1-\frac{1}{2}\frac{2ic+c^2}{r_{ij}^2}+\frac{3}{8}\left(\frac{2ic+c^2}{r_{ij}^2}\right)^2\right)$$

$$=\frac{1}{r_{ij}}\left(1-\frac{1}{2}\frac{2ic+c^2}{r_{ij}^2}+\frac{3}{2}\frac{i^2c^2}{r_{ij}^4}\right) \quad r_{ij}=\sqrt{i^2+j^2} \quad (2)$$

And the lattice energy is calculated as:

$$E=\frac{e^2}{4\pi\varepsilon}\sum_{i,j}\left\{\frac{P_{Cu}^2}{r_{ij}}+\frac{2P_O^2}{r_{ij}}+\frac{P_O^2}{r_{ij}}\left[2+\frac{c^2}{r_{ij}^2}\right]+\frac{P_{Cu}P_O}{r_{ij}}\left[4+\frac{c^2}{r_{ij}^2}\right]\right\}$$

$$=\frac{e^2}{4\pi\varepsilon}(P_{Cu}+2P_O)^2\sum_{i,j}\frac{1}{r_{ij}}+\frac{e^2}{4\pi\varepsilon_0}(P_O^2+P_{Cu}P_O)\sum_{i,j}\frac{c^2}{r_{ij}^3} \quad (3)$$

$$=\frac{e^2}{4\pi\varepsilon}\sum_{i,j}\frac{1}{r_{ij}}+\frac{e^2}{4\pi\varepsilon}\frac{1-P_{Cu}^2}{4}\sum_{i,j}\frac{c^2}{r_{ij}^3}$$

Where $e$ is charge of carriers $i,j$ represents the position of different units on the lattice. Take into account the actual $a_0$, lattice energy is written as:

$$E=E_0+E_{dipole} \quad E_0=\frac{e^2}{4\pi\varepsilon a_0}\sum_{i,j}\frac{1}{r_{ij}} \quad E_{dipole}=\frac{e^2\cdot b^2}{4\pi\varepsilon a_0^3}\frac{1-P_{Cu}^2}{4}\sum_{i,j}\frac{1}{r_{ij}^3} \quad (4)$$

The first term $E_0$ comes from the regular ion lattice, means that all ion charge being on Cu ions, and it keeps electricity equilibrium with carriers; the second term $E_{dipole}$ comes from the position difference between Cu-O ions, means that ion charge being on O ions instead of Cu ions, and it violates the electricity equilibrium slightly. Since the $E_{dipole}$ acts like the interactivity between electronic dipoles, I name it as electronic dipole lattice, and the ion lattice on $CuO_2$ plane can be seen as combination of regular lattice and dipole lattice. One can see that $E_{dipole}$ is far less than $E_0$, means that the approximation of only keeping $c^2$ term around optimum doping is suitable. And one should also pay attention to the fact that equation (4) is only founded for small $x_0$, higher order terms should be considered for bigger $x_0$.

Equation (4) can be seen as the proof of the existence of electronic dipoles lattice. The electronic dipoles lattice comes from the position difference of Cu-O ions and low carrier concentration, is the innate character of $CuO_2$ plane. The interactivity between regular lattice and carriers acts like normal metal forms the ground state; the electronic dipole lattice induces additional energy to the $CuO_2$ plane, dominates the character of cuprate superconductor.

## 3 Explanation of cuprate superconductor phenomena

In consideration of electronic dipole lattice, the special character of cuprate superconductor is tersely interpreted, and can be seen as the proof of the existence of electronic dipole lattice.

### 3.1 Paring mechanism

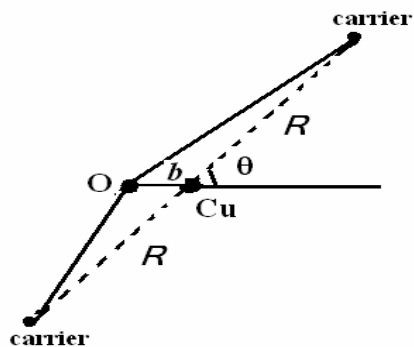

**Fig2. Sketch of the electronic dipole**

Fig.2 illustrates the creation of electronic dipole lattice, where $2R$ is coherence length of paired carriers, and is dominated by cuprate sample. Experimental data shows that bond length of Cu-O $b$ is far less than $R$ [11]. Since the low concentration of charged ions around optimum doping, the image method that well known in classical electrodynamics is applicable. When the charge is on O ion instead of Cu ion, it acts like an electronic dipole, and these electronic dipoles form the electronic dipole lattice. Assuming pairing occurs in real space, the additional energy between two carriers opposite of a dipole is:

$$V_e = \frac{1}{\pi} \frac{e^2}{4\pi\varepsilon} \int_0^\pi \left( \frac{2}{R} - \frac{1}{\sqrt{(R\cos\theta - b)^2 + R^2\sin^2\theta}} - \frac{1}{\sqrt{(R\cos\theta + b)^2 + R^2\sin^2\theta}} \right) d\theta \approx -\frac{e^2}{8\pi\varepsilon} \frac{b^2}{R^3} \quad (5)$$

This simple model shows that the additional energy between two carriers induced by dipole lattice can be negative, and according to BCS theory, pairing happens in such situation. From equation (4) one can find that the concentration of electronic dipoles $N_{dipole}$ is about:

$$N_{dipole} = (1 - P_{Cu}^2)^{1/2} / 2 * x_0 \quad (6)$$

It shows that $N_{dipole}$ is less than $x_0/2$, and the condition that two carriers paired through an electronic dipole is satisfied.

### 3.2 The bell-shaped phase diagram

The energy gap $\Delta$ of cooper pairs is determined by the additional energy induced by dipole lattice, which includes the energy between paired carriers and that between electronic dipoles. Considering the number of electronic dipoles $N_{dipole}$, when only take account of the nearest interactivity of electronic dipoles, the energy gap can be approximately written as:

$$\Delta \approx \left| \frac{\sqrt{1-P_{Cu}^2}}{2} V_e + E_{dipole}/2 \right| = \left| -\frac{e^2}{8\pi\varepsilon_0} \frac{\sqrt{1-P_{Cu}^2}}{2} \frac{b^2}{R^3} + \frac{1-P_{Cu}^2}{4} \frac{e^2}{2\pi\varepsilon_0} \frac{b^2}{a_0^3} \right| \quad (7)$$

$$= \frac{e^2}{8\pi\varepsilon_0} \frac{\sqrt{1-P_{Cu}^2}}{2} \frac{b^2}{R^3} \left(1 - \alpha \cdot x_0^{3/2}\right)$$

Where $\alpha = 2\sqrt{1-P_{Cu}^2} \frac{R^3}{b^3}$. It shows that when $P_{Cu}=1$ the energy gap disappears, and when $P_{Cu}<1$ the energy gap always exists.

The concentration of cooper pairs $N_{pairs}$ is proportional to $N_{dipole}$, and the breaking probability of cooper pairs increases with enhancement of temperature $T$. According to the Boatman distribution, the probability of carrier energy being less than energy gap $\Delta$, that means cooper pairs keep unbroken, is $\int_0^\Delta \exp(-\Delta/k_B T) / \int_0^\infty \exp(-\Delta/k_B T) = 1 - \exp(-\Delta/k_B T)$, and the concentration of Cooper pairs $N_{pairs}$ is:

$$N_{pairs} \approx N_{dipole}\left[1-\exp(-\Delta/k_B T)\right] = 0.5 * (1-P_{Cu}^2)^{1/2} x_0 \left[1-\exp(-\Delta/k_B T)\right] \quad (8)$$

Since the low carrier concentration, the transition temperature $T_c$ is influenced by percolation model [10], there exist a critical concentration of cooper pairs $N_c$, only when the condition of equation (9) is satisfied, superconductivity occurs:

$$N_{pair} \geq N_c \quad (9)$$

The critical transition temperature $T_c$ can be calculated by $N_{pair}=N_c$, and is dominated by three factors $P_{Cu}$, $x_0$, and $\Delta$.

Since $P_{Cu}$ is often bigger than 0.5, and the maximum critical transition temperature $T_{c\max}$ usually occurs at $P_{Cu} = 0.5$ [12], when $x_0$ is about 0.15. It is known that energy gap of pairing carriers is about $3k_BT - 4k_BT_c$ [13], take $\Delta = 4k_BT_c$, one can get $N_c = 0.063$.

From equation (8) and equation (9), one can get

$$T_c \propto \frac{-x_0^{3/2}}{\log\left(1 - \frac{0.063}{x_0}\right)} \quad (10)$$

Since $P_{Cu}$ and $\alpha$ vary with the variation of $x_0$, to quantitatively determine the constants, detailed model considering the material should be considered.

It can be easily found that the curve determined by Equation (10) is bell-shaped. For example, in fig.3 the curves of normalized $T_c/T_{c\max} = (1 - 7 \cdot x_0^{3/2})/\log(1 - 0.05/x_0)$ and the empirical relation $\frac{T_c}{T_{c\max}} = 1 - 82.6(x_0 - 0.16)^2$ [14] are compared.

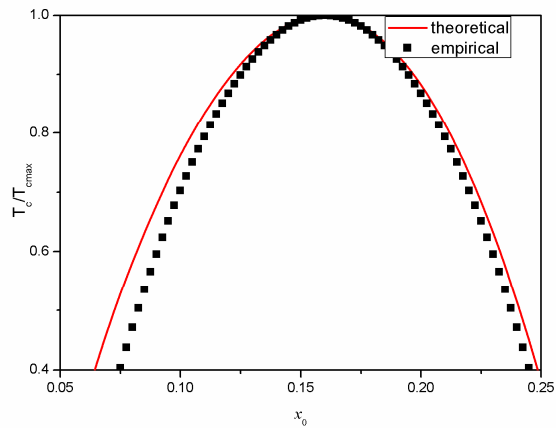

**Fig3. compare of phase diagram between theoretical and empirical curve**

### 3.3 Pseudogap

Considering about equation(7), (8) and (9), one can find that, in the underdoped situation, $E_{dipole}$ is relatively small, energy gap almost keeps constant and is relatively big; in the overdoped situation, length between ions is relatively small, $E_{dipole}$ can't be omitted, and it increases with the enhancement of carrier concentration, so energy gap $\Delta$ decreases with the enhancement of $x_0$.

In the underdoped situation, $\Delta$ is almost constant and is big enough, so $T_c$ increases with the increase of $x_0$, and this also illustrates the existence of pseudogap. In the overdoped situation, $x_0$ is big enough and its effect can't be omitted. Since $T_c$ decreases with the decrease of $\Delta$, and $\Delta$ decreases with the enhancement of $x_0$, so $T_c$ decreases with the enhancement of $x_0$.

From equation (8) and equation (9), one can find that for same $x_0$, $T_c$ increases with the decrease of $P_{Cu}$, and this conclusion is supported by the study of Guo-Qing Zheng et. al. [12].

### 3.4 d-wave symmetry

Since the dipole only creates on Cu-O bond direction, the d-wave symmetry naturally occurs, one can find from equation (5) and equation (7).

### 3.5 Linear temperature dependence of normal state resistivity

Since ion lattice on $CuO_2$ plane is combination of regular lattice and dipole lattice, the resistivity of normal state comes from the dispersion of both these two lattices. On the other hand, as the dipole lattice creates from the violation of regular lattice, they should have similar physical character.

When the temperature is just above $T_c$, the characteristic energy of regular lattice is about $k_B T_c$, and the characteristic energy of dipole lattice is about the energy gap of pairing carriers which is about $3k_B T - 4k_B T_c$ [13], so the effective amplitude of electronic dipole lattice is much larger than that of regular lattice, and resistivity is mainly dominated by electronic dipole lattice.

Considering the relationship between Debye temperature $T_d$ and lattice number of 2D lattice [15], one get the relationship $T_d \propto N_{dipole}^{1/2}$, and the ratio of Debye temperature between different lattices is

$$\frac{T_{d-dipole}}{T_{d-regular}} = \frac{(1-P_{Cu}^2)^{1/2}}{2} \quad (11)$$

Since $P_{Cu}$ is less than 1, the Debye temperature of dipole lattice $T_{d-dipole}$ is far less than that of regular lattice $T_{d-regular}$. As $T_{d-dipole}$ is relatively little, the condition $T_c > T_{d-dipole}/3$ will be easily satisfied, and the normal state resistivity follows the linear temperature dependence [16].

## 4 Conclusions

In consideration of position difference of Cu-O ions, it is proved that the ion lattice on CuO$_2$ plane can be seen as combination of regular lattice and electronic dipole lattice, and the electronic dipole lattice is the innate character of CuO$_2$ plane. The existence of electronic dipole lattice predicts the character of cuprate superconductor, such as the bell-shaped phase diagram, pseudogap, d-wave symmetry, and the linear temperature dependence of normal state resistivity.

The concept of hole comes from band theory, and band theory bases on the period structure. So the normal concept of hole neglects the position difference of Cu-O ions, and can't explain the character of cuprate superconductor thoroughly.

In conclusion, the existence of electronic dipole lattice is supported by theoretical analysis and experiment phenomena explanation, and it maybe an important factor that affect the character of cuprate superconductor, and worth for further study.

11237 (1990)